# Formation mechanisms of magnetic bubbles in an *M*-type hexaferrite: the role of chirality reversals at domain walls

H. Nakajima[1], A. Kotani[1], K. Harada[1, 2], Y. Ishii[1], and S. Mori[1]

[1]*Department of Materials Science, Osaka Prefecture University, 599-8531, Japan*

[2]*Center for Emergent Matter Science, the Institute of Physical and Chemical Research (RIKEN),*

*Hatoyama, Saitama 350-0395, Japan*

## ABSTRACT

We examined the formation mechanisms of magnetic bubbles in an *M*-type hexaferrite via Lorentz microscopy. When magnetic fields were perpendicularly applied to a thin sample of $BaFe_{12-x-0.05}Sc_xMg_{0.05}O_{19}$ ($x = 1.6$), Bloch lines, which were identified as reversals of domain-wall chirality, appeared, and magnetic bubbles were formed when the magnetic stripes were pinched off at these Bloch lines. The number of Bloch lines increased with the amount of Sc in $BaFe_{12-x-0.05}Sc_xMg_{0.05}O_{19}$ probably because of the reduction in magnetic anisotropy. A Lorentz microscopic observation revealed that Bloch lines with high magnetostatic energy may play an important role in the formation of magnetic bubbles.

## I. INTRODUCTION

Nanoscale magnetic spin textures such as skyrmions and magnetic bubbles have attracted significant interest because of their novel physical phenomena and potential applications in spintronics and high-density magnetic storage devices.[1] Skyrmions are vortex-like swirling spin textures induced by the Dzyaloshinskii–Moriya interaction in chiral magnets; these textures exhibit the topological Hall effect and ultralow current-driven motion.[2,3] Conversely, magnetic bubbles are cylindrical domains stemming from magnetocrystalline anisotropy in uniaxial magnetic thin films.[4, 5] Magnetic bubbles are formed by a subtle balance among the electrostatic, domain wall, and Zeeman energies. Magnetic bubbles originating from the magnetic dipole–dipole interaction in uniaxial magnets have a circularly rotating spin structure within a domain wall and are considered to have the same topological skyrmion number ($S = 1$) as that of skyrmions in chiral magnets; this allows a current-driven motion to be realized in skyrmions and magnetic bubbles.[1,7-9] Thus, magnetic bubbles, including other nanometric vortices, have recently attracted renewed interest.[9-14]

In addition to these studies, Lorentz microscopy revealed that magnetic bubbles in $M$-type hexaferrites BaFe$_{12-x-0.05}$Sc$_x$Mg$_{0.05}$O$_{19}$ ($x = 1.6$, 1.8) exhibit unusual physical phenomena such as multiple-helicity and thermally activated helicity reversals.[15,16] In the previous study using Lorentz microscopy[15], magnetic stripes changed into magnetic bubbles with a multi-ring structure under the application of magnetic fields perpendicular to a thin film, with the bubbles ultimately shrinking and disappearing above a critical value of ~200 mT. Although the appearance of magnetic bubbles with increasing strength of the vertical magnetic field as well as the bubbles' spin configurations have been investigated, the formation mechanisms of such bubbles in BaFe$_{12-x-0.05}$Sc$_x$Mg$_{0.05}$O$_{19}$ are yet to be revealed. In addition, the effect of partial substitution of Sc for Fe in BaFe$_{12-x-0.05}$Sc$_x$Mg$_{0.05}$O$_{19}$ remains elusive.

In this study, we investigated the formation mechanisms of magnetic bubbles in a Sc-substituted $M$-type hexaferrite, BaFe$_{12-x-0.05}$Sc$_x$Mg$_{0.05}$O$_{19}$, using *in situ* Lorentz microscopy. We successfully observed that Bloch lines, which are chirality reversals of Bloch walls, are formed under the application of vertical magnetic fields. Furthermore, magnetic bubbles are formed by pinching off stripe domain walls at Bloch lines. Compared with the results for a hexaferrite without substitution, the magnetic anisotropy of a Sc-substituted hexaferrite is smaller; this is attributed to the appearance of Bloch lines and the formation mechanism of the magnetic bubbles.

## II. MATERIALS AND METHODS

Figure 1(a) shows the crystalline structure of an $M$-type hexaferrite and a ferrimagnetic spin structure composed of two blocks ($R$ and $R$'). The net magnetization points along the $c$ axis in the hexagonal system;

however, the spins parallel and anti-parallel to the $c$ axis coexist in each $R$ and $R$'.[17] The Sc-substituted $M$-type hexaferrite shows ferrimagnetism at room temperature and longitudinal spiral magnetism below 270 K in BaFe$_{12-x-0.05}$Sc$_x$Mg$_{0.05}$O$_{19}$ ($x = 1.6$), thereby resulting in spontaneous electric polarization induced by external magnetic fields.[18]

Single crystals of the $M$-type hexaferrites BaFe$_{12-x-0.05}$Sc$_x$Mg$_{0.05}$O$_{19}$ ($x = 1.6$) and BaFe$_{12}$O$_{19}$ were grown using a floating zone method in oxygen gas flow.[18] The crystals were cut into thin plates with wide surfaces perpendicular to the crystallographic $c$ axis and thinned by mechanical grinding and Ar$^+$ ion milling. Observations were made at room temperature using transmission electron microscopes operating at 200 kV (JEM-2010 and JEM-2100F). External magnetic fields were applied with the objective lens perpendicular to the thin film or parallel to the $c$ axis. The Fresnel method of Lorentz microscopy was utilized to observe magnetic domain walls. In the method, domain walls are visualized as bright and/or dark contrasts owing to the Lorentz force in defocused condition.[19, 20]

## III. RESULTS AND DISCUSSION

First, we observed stripe domain structures under the no-magnetic-field condition. Figure 2(a) shows a Fresnel image of the Sc-substituted $M$-type hexaferrite along the $c$ axis. Magnetic domain walls are visualized as pairs of bright and dark contrasts due to the Lorentz deflection; this indicates that the magnetic domains have their magnetization parallel to the $c$ axis and that the Bloch walls separate the up-magnetized and down-magnetized domains. The magnetic domain structure has a periodic undulating character, which results from the competition between the magnetostatic and magnetocrystalline anisotropic energies. Note that the broad, curved black lines are bend contours. Next, we applied a magnetic field of 100 mT perpendicular to the thin film, as shown in Fig. 2(b). The widths of the up-magnetized domains decreased because the external magnetic field was applied along the downward direction. Figure 2(c) shows a magnified image and line profiles of the intensity from a region marked by a rectangle in Fig. 2(b). The line profile of the intensity along 1-2 shows contrasts of the walls comprise dark-bright and bright-dark lines. Meanwhile, two dark-bright lines are observed along 3-4, demonstrating the magnetic domain wall's chirality is reversed at the right domain wall. The reversal of the domain wall chirality is called a Bloch line and the directions of the in-plane magnetization are reversed by rotating gradually within a domain wall at a Bloch line, as shown in Fig. 1(b). By dividing a domain wall into some segments, the stray field energy stemming from the wall is considered to be reduced.[21] As can be seen in Fig. 2(b), a lot of Bloch lines, which could be observed as reversals in a sequence of bright and dark contrasts, were created under an external magnetic fields. Conversely, only a few Bloch lines were observed without the application of a magnetic field, as shown in Fig. 2(a).

To investigate a formation mechanism of magnetic bubbles in $BaFe_{12-x-0.05}Sc_xMg_{0.05}O_{19}$ ($x = 1.6$), we observed the magnetic-field dependence of the magnetic domain walls. Figure 3 shows Fresnel images of the magnetic domain walls under fields ranging from 0 to 204 mT. In Fig. 3(a), stripe magnetic domain walls, similar to those shown in Fig. 2(a), are observed without an external magnetic field. The magnetic domains with up- and down-magnetizations were equally populated, and they were separated from one another at the Bloch walls. When a magnetic field was applied perpendicular to the plane, the magnetic domains with antiparallel magnetization shrank, as shown in Fig. 3(b). Further increasing the magnetic field decreased the area of up-magnetized domains [Fig. 3(e)], and a Bloch line was observed, as indicated by a blue arrow. A schematic of the magnetic domain structure around the Bloch line is shown in Fig. 3(i). At 195 mT, as shown in Fig. 3(d), some domain walls changed into clockwise, counterclockwise type-I, and type-II bubbles. Note that type-I magnetic bubbles are defined as magnetic domain walls, which are continuously magnetized clockwise or counterclockwise, whereas type-II bubbles are defined as opposed magnetizations in two halves of the wall producing a pair of Bloch lines.[22] Their schematics and the corresponding Fresnel images are shown in Fig. 1(c). At the same time as the formation of the magnetic bubbles, the Bloch line moved toward the edge of the Bloch wall. The magnetic domain was isolated at 200 mT [Fig. 3(e)]. The transition from stripe magnetic domain walls to magnetic bubbles is first-order;

thus, the two structures can coexist in a range of magnetic fields. The edges were expanded because the repulsion of the magnetic domain wall and Bloch lines were generated, as indicated by arrowheads. In Fig. 3(f), the Bloch wall was pinched off at the Bloch line by increasing the magnetic field by 2 mT. Comparing Figs. 3(e) and 3(f), it is found that the lower side of the domain wall changed into a counterclockwise type-I bubble, labeled as α. The counterclockwise magnetic bubble α was formed from the Bloch walls whose magnetizations were pointed parallel to each other, as indicated by red arrows; thus, the chirality was randomly determined. The upper side of the magnetic domain shrank to form a clockwise type-I magnetic bubble β [Fig. 3(f), 3(g), and 3(h)]. Chirality was conserved during the formation of the magnetic bubble β, and the two Bloch lines at the edge were annihilated. After the stripe magnetic walls changed into magnetic bubbles, a mixture of type-I and type-II bubbles moved to form a regular triangular lattice, as shown in Figs. 3(g) and 3(h).

Here let us discuss the formation mechanism of magnetic bubbles. A formation mechanism of type-II magnetic bubbles in a Co thin film was proposed by Grundy *et al.*[22] According to them, stripe magnetic domain walls orient toward the in-plane component of an external magnetic field when the magnetic easy axis is tilted away from the field. By increasing the magnetic fields, Bloch lines are created at the edges of Bloch walls, thereby pinching off the magnetic walls; the Bloch walls shrink to become type-II magnetic bubbles. Figure 4 shows Fresnel images in the range of 135–200 mT, with the *c* axis of the sample tilted by 5° from the magnetic fields in the *M*-type hexaferrite. Contrary to that observed in Fig. 2(b), Bloch lines were rarely observed in Fig. 4, except in a few curved regions. The Bloch walls were oriented along the same direction as the magnetic field because of the Zeeman energy. Comparing Figs. 4(b) and 4(c), Bloch walls were pinched off at the positions indicated by blue arrows, and Bloch lines were created at the edges of the magnetic domain walls. Then, the pinched domain walls shrank to become type-II magnetic bubbles maintaining the positions of the Bloch lines, as shown by yellow arrows. Note that the observed magnetic bubbles are all type-II. The observed results in Fig. 3 are totally different from those observed in Fig. 4 and cannot be explained in terms of the pinching process as observed in Fig. 4 and the study of a Co thin film[22] because the created bubbles are type-I and the in-plane component of the magnetic field is small. We considered another mechanism to explain the observed behaviors, which is described as follows. Bloch lines are generated to reduce the magnetostatic energy originating from the Bloch walls;[21] however, a Bloch line itself has energies higher than a Bloch wall by 7.9 erg/cm$^2$ because the magnetizations in a Bloch line are not parallel, which reduces the energy gain of ferromagnetic exchange interactions [See Fig. 2(b)].[22, 23] Therefore, when a magnetic bubble is created, pinching off the stripe at which a Bloch line is located makes the system more stable.

The formation mechanism observed in this study is characteristic of a uniaxial magnet whose magnetic anisotropy is reduced by the substitution of Sc for Fe. To clarify the effect of this substitution, we observed an $M$-type hexaferrite, $BaFe_{12}O_{19}$, without substitution under external magnetic fields. Figure 5(a) and (b) show Fresnel images of the Sc-substituted and Sc-free hexaferrites. We observed many Bloch lines with magnetic bubbles in Fig. 5(a) as the magnetic field increased. Conversely, no Bloch lines were seen in Fig. 5(b), except at the edges of the magnetic domains in $BaFe_{12}O_{19}$. A Bloch line has its magnetization components perpendicular to the $c$ axis and is thus required to reduce the magnetic anisotropy to form Bloch lines; this explains the results observed in Fig. 5. We found that the diameter of the magnetic bubbles in Fig. 5(a) (~180 nm) is smaller than that in Fig. 5(b) (~360 nm). This can be attributed to the reduction in the magnetocrystalline anisotropy.[24]

The Sc-substituted hexaferrite was also confirmed to have a smaller magnetic anisotropy by magnetization measurements. Figure 5(c) shows the magnetic-field dependence of the magnetization at room temperature for $BaFe_{12}O_{19}$ and $BaFe_{12-x-0.05}Sc_xMg_{0.05}O_{19}$ ($x = 1.6$). The uniaxial anisotropy coefficient $K_u$ is expressed by $K_u = H_kM_s/2$, where $H_k$ is the anisotropy field, which is defined as the magnetic field in which the easy axis and the hard axis of the magnetization have the same value (as indicated by arrowheads in Fig. 5(c)), and $M_s$ is the saturation magnetization.[21] By substituting Sc for Fe, $H_k$ decreases from 4.5 T to 0.37 T, and $K_u$ is reduced from $8.4 \times 10^5$ J/m$^3$ to $1.0 \times 10^5$ J/m$^3$. Hence, the Sc-substituted hexaferrite is found to have a magnetocrystalline anisotropy smaller than that obtained without substitution by approximately 90%.

## IV. CONCLUSIONS

We directly observed the formation mechanisms of magnetic bubbles in Sc-substituted *M*-type hexaferrite using Lorenz microscopy. Bloch lines were created when external magnetic fields were applied perpendicularly to a thin film. Moreover, a magnetic bubble was created by pinching off a Bloch line; this is due to the instability of the Bloch line in terms of energy. We have shown that a reduction in magnetic anisotropy is essential for this mechanism to occur. These results represent important information concerning the relationship between Bloch lines and magnetic bubbles and are expected to encourage applications of magnetic vortices, such as magnetic bubbles and skyrmions.

## ACKNOWLEDGMENTS

This work was partially supported by Grant-in-Aid (No. 16H03833 and No. 15K13306) from the Ministry of Education, Culture, Sports, Science and Technology (MEXT), Japan.

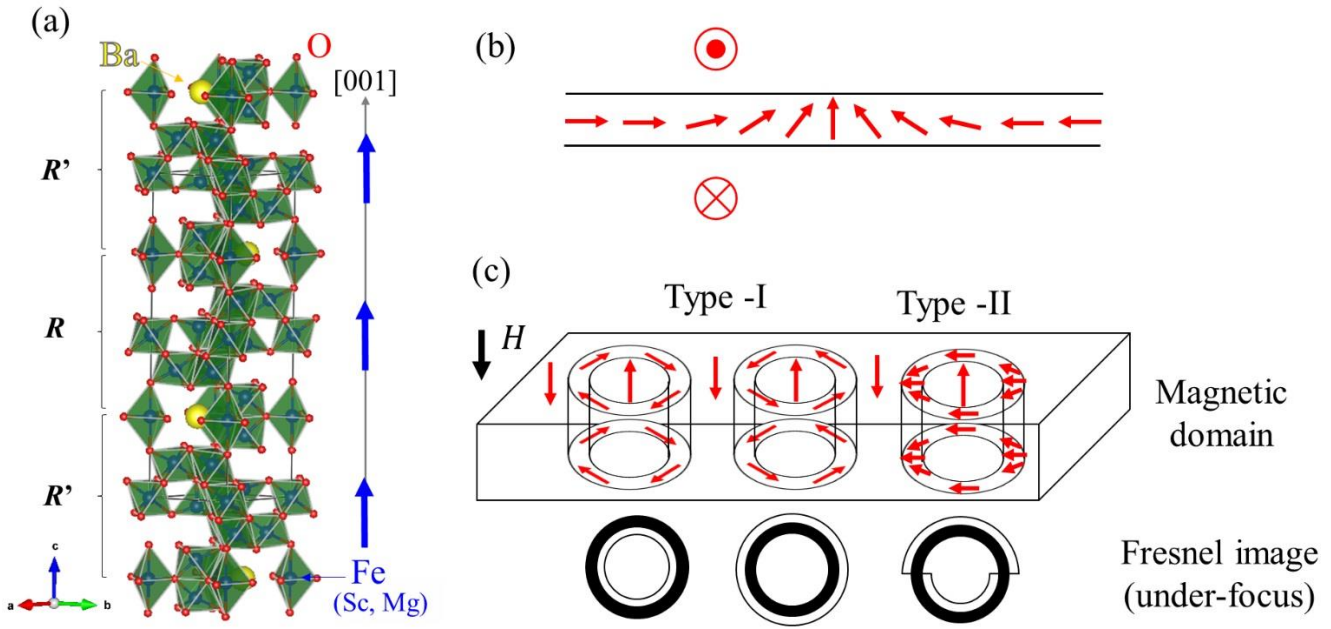

FIG. 1. (a) Crystalline and magnetic structures of an *M*-type hexaferrite. Blue arrows represent magnetic moments at each block (R and R'). (b) Spatial configuration of magnetization of a Bloch line in a domain wall. (c) Schematic illustrations of type-I and -II magnetic bubbles and the corresponding contrasts of Fresnel images (under-focus condition). Red arrows represent the magnetization.

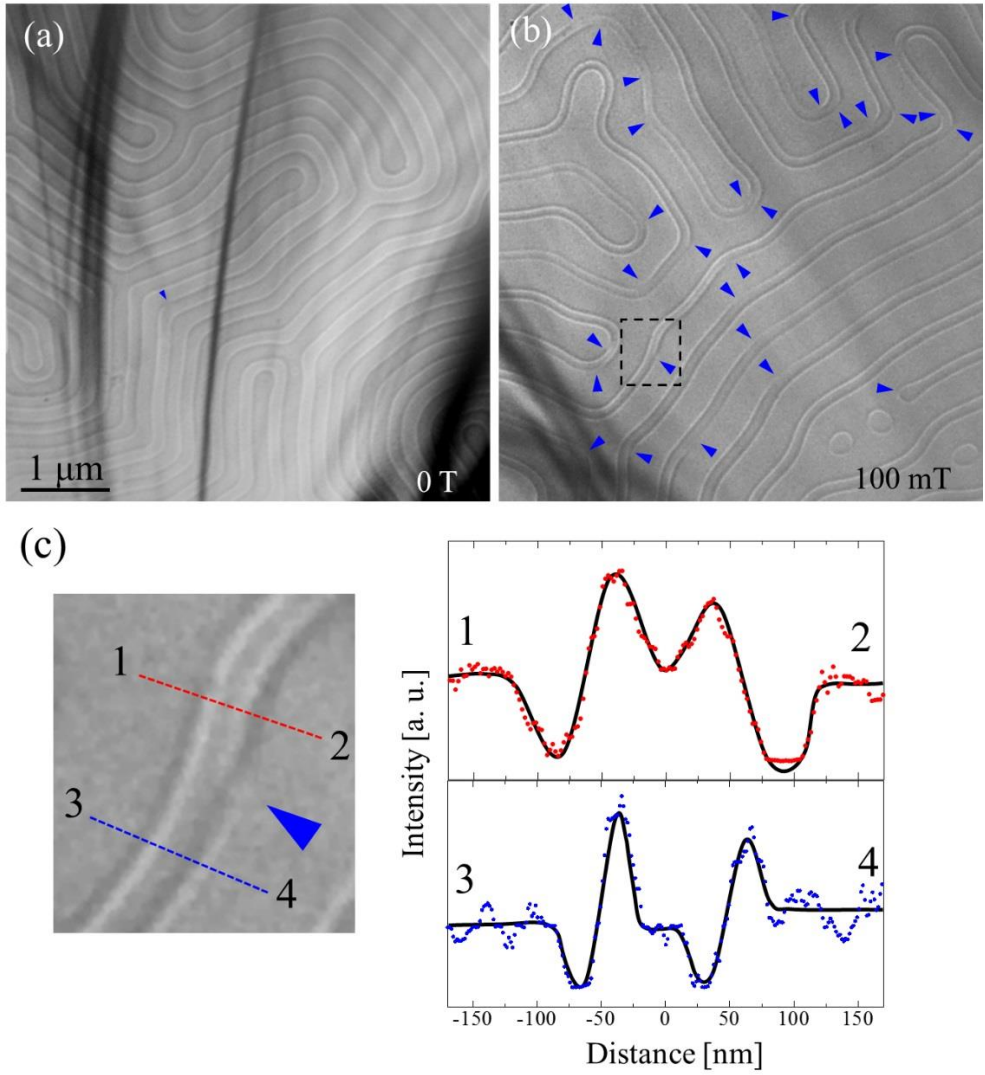

FIG. 2. Fresnel images (a) without magnetic fields and (b) at 100 mT. The defocus value is −200 nm (under-focus). Blue arrowheads indicate Bloch lines. (c) Line profiles of the intensity along 1-2 and 3-4 around a Bloch line marked by a rectangle in (b). In the profiles, solid lines are provided to guide the eye.

.

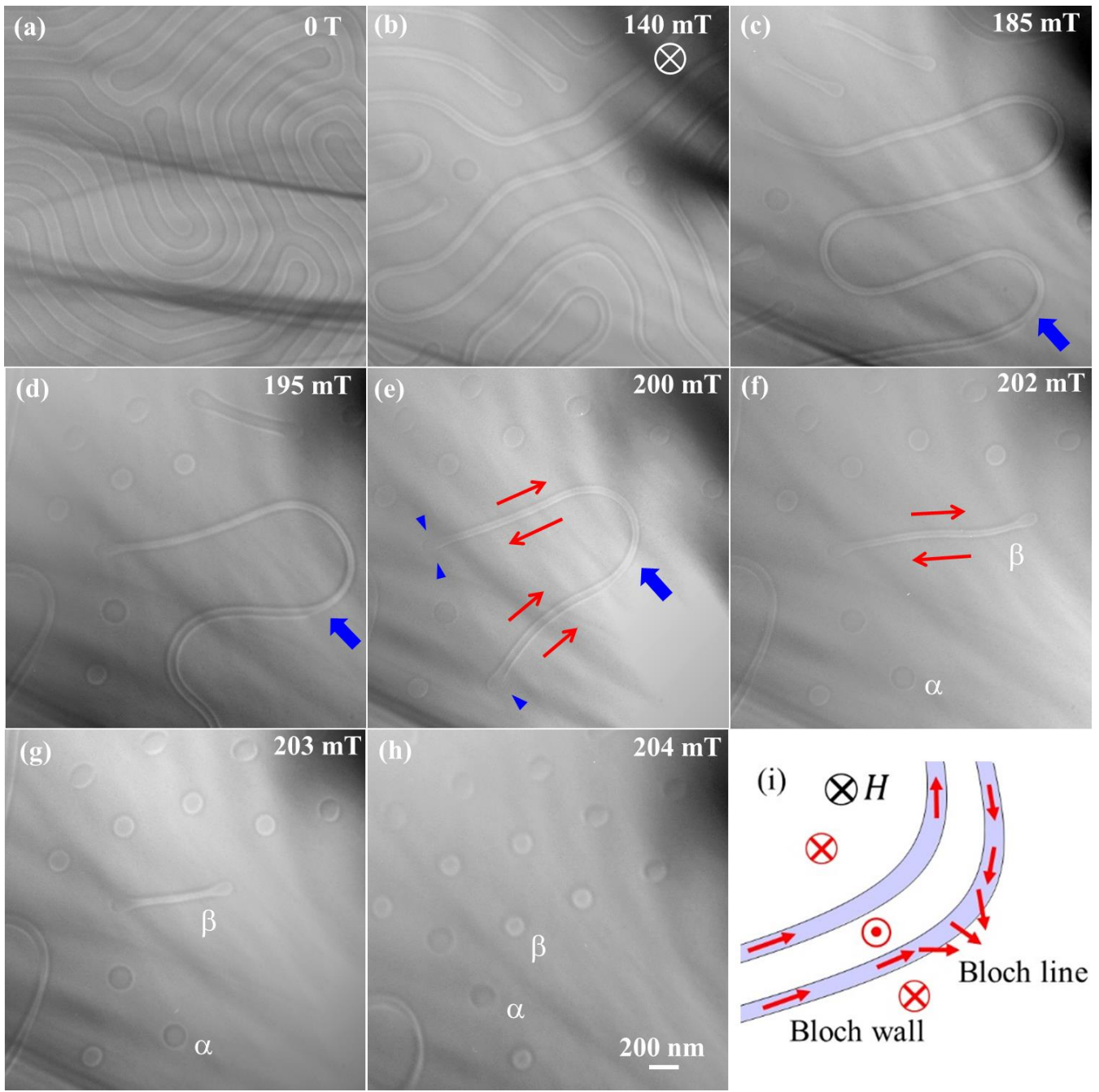

FIG. 3. (a)–(h) Fresnel images at magnetic fields of 0–204 mT. (i) Schematic of the Bloch walls and Bloch line shown in Fig. 3(e). Red arrows represent the magnetizations of the Bloch walls. Blue arrowheads represent Bloch lines. The same domains are labeled as α and β in (f)–(h).

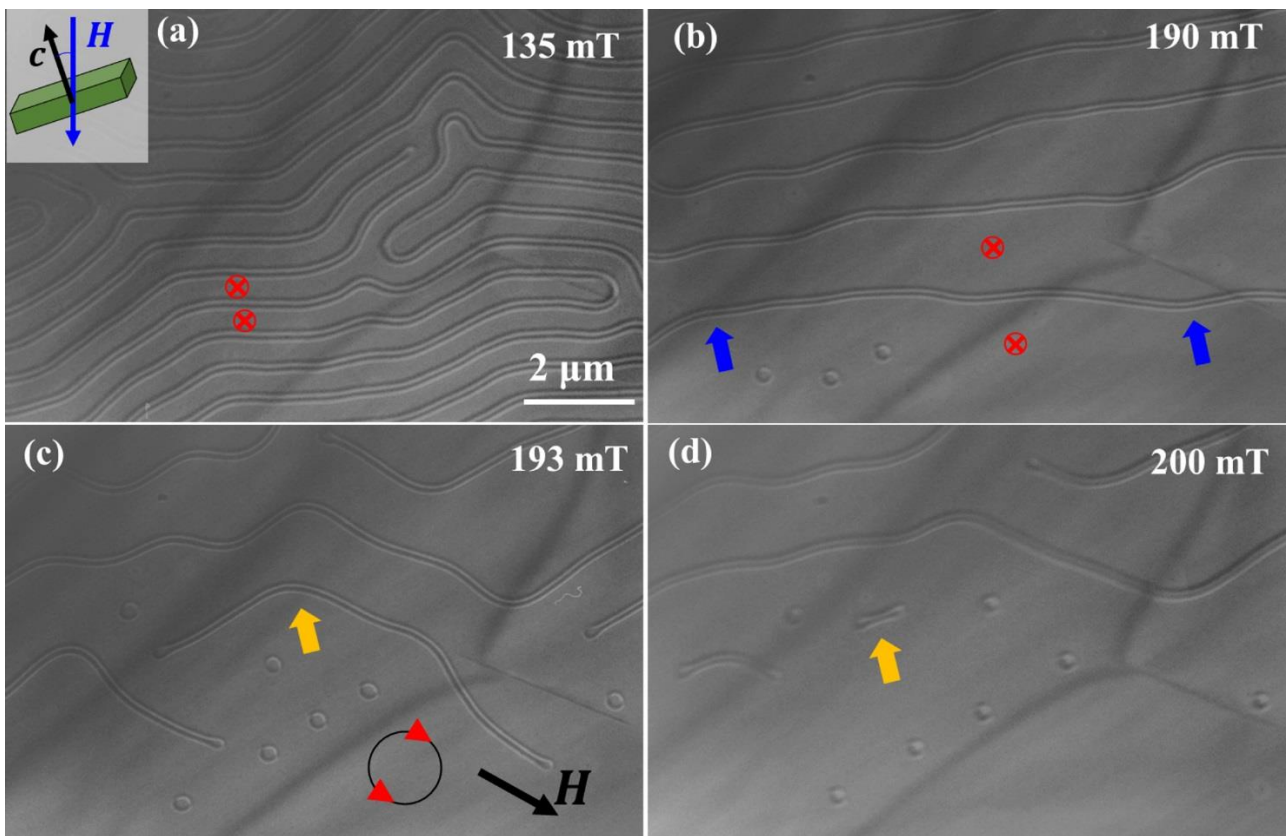

FIG. 4 Fresnel images with increasing magnetic fields. The specimen was tilted by 5° from the magnetic field, as shown in the inset. Blue arrows in (b) show the points at which the stripe was pinched off in (c).

.

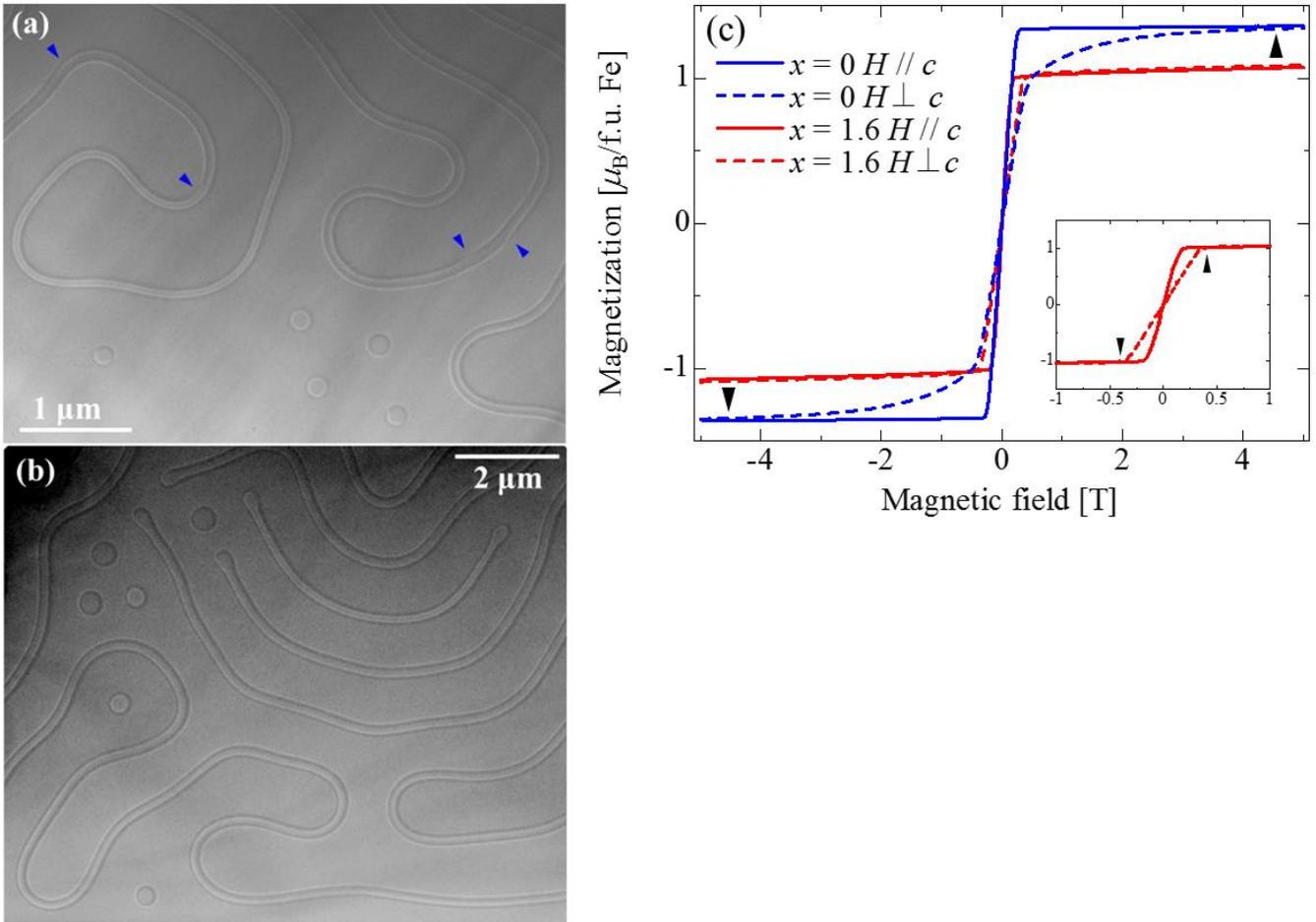

FIG. 5 Under-focused Fresnel images of (a) Sc-substituted and (b) Sc-free (BaFe$_{12}$O$_{19}$) $M$-type hexaferrites under a magnetic field of 180 mT. Blue arrowheads show Bloch lines. (c) Magnetic-field dependence of the magnetization at room temperature for $x = 0$ (BaFe$_{12}$O$_{19}$) and $x = 1.6$ (BaFe$_{12-x-0.05}$Sc$_x$Mg$_{0.05}$O$_{19}$). The anisotropy fields $H_k$ are indicated by arrowheads.